\documentclass[12pt,preprint]{aastex}
\usepackage{graphicx}
\usepackage{epstopdf}
\usepackage{rotating}
\usepackage{float}
\shorttitle{Flux Anomalies in SDSS J1029+2623}
\shortauthors{Kratzer et al.}

\begin{document}

\title{Analyzing the Flux Anomalies of the Large-Separation Lensed Quasar SDSS J1029+2623}
\author{Rachael M. Kratzer,\altaffilmark{1} Gordon T. Richards,\altaffilmark{1} David M. Goldberg,\altaffilmark{1} Masamune Oguri,\altaffilmark{2} Christopher S. Kochanek,\altaffilmark{3}  Jacqueline A. Hodge,\altaffilmark{4}  Robert H. Becker,\altaffilmark{4,5} Naohisa Inada\altaffilmark{6}}

\altaffiltext{1}{Department of Physics, Drexel University, Philadelphia, PA.}
\altaffiltext{2}{Division of Theoretical Astronomy, National Astronomical Observatory of Japan, 2-21-1 Osawa, Mitaka, Tokyo 181-8588, Japan.}
\altaffiltext{3}{Department of Astronomy, The Ohio State University, Columbus, OH.}
\altaffiltext{4}{Department of Physics, University of California, Davis, CA.}
\altaffiltext{5}{Institute for Geophysics and Planetary Physics, Lawrence Livermore National Laboratory.}
\altaffiltext{6}{Research Center for the Early Universe, School of Science, University of Tokyo, Bunkyo-ku, Tokyo 113-0033, Japan.}

\begin{abstract}
Using a high resolution radio image, we successfully resolve the two fold image components B and C of the quasar lens system SDSS J1029+2623. The flux anomalies associated with these two components in the optical regime persist, albeit less strongly, in our radio observations suggesting that the cluster must be modeled by something more than a single central potential. We argue that placing substructure close to one of the components can account for a flux anomaly with negligible changes in the component positions. Our best fit model has a substructure mass of $\sim 10^{9}$ M$_\sun$ up to the mass-sheet degeneracy, located roughly $0\farcs1$ West and $0\farcs1$ North of component B. We demonstrate that a positional offset between the centers of the source components can explain the differences between the optical and radio flux ratios.
\end{abstract}


\section{Introduction}
SDSS J102913.94+262317.9 (SDSS J1029+2623; Inada et al. 2006) is only the second ``naked cusp'' lens  after APM 08279+5255 (Lewis et al. 2002) to be discovered, and with a 22\farcs6 maximum separation between its components, it is the largest known quasar lens. These attributes alone would make it an interesting lens system, but it also exhibits one of the more dramatic examples of an anomalous flux ratio which cannot be reproduced by a single central potential. Such anomalies can be produced by differential extinction (Lawrence et al. 1995; Falco et al. 1999; El\'{i}asd\'{o}ttir et al. 2006), microlensing by stars in the lens (Koopmans \& de Bruyn 2000; Morgan et al. 2006; Poindexter et al. 2007; Anguita et al. 2008) or the presence of substructure (satellites) in the primary lens or along the line of site (Mao \& Schneider 1998; Ros et al. 2000; Metcalf \& Madau 2001; Dalal \& Kochanek 2002; Kochanek \& Dalal 2004; MacLeod et al. 2009). The latter case is of particular interest with regard to the ``missing satellite" problem (Klypin et al. 1999; Moore et al. 1999; Brada\u{c} et al. 2002; Chiba et al. 2005; Miranda \& Jetzer 2007).

Originally discovered during the Sloan Digital Sky Survey Quasar Lens Search (SQLS; Oguri et al. 2006, 2008a; Inada et al. 2008), Inada et al. (2006) spectroscopically confirmed that the A and B components of  SDSS J1029+2623 (see Figure 1) captured by the Sloan Digital Sky Survey (SDSS; York et al. 2000) were lensed images of the same radio-loud quasar located at $z_{s} = 2.197$. A mass model based on these two images predicted a potential that was inconsistent with the location of the observed lens galaxy/cluster, necessitating further investigation. Follow-up optical and spectral observations made by Oguri et al. (2008b) resolved the B component into two components, B and C, and the lensing galaxy, G1, into two components (G1a and G1b). Additional spectra confirmed that C was indeed another lensed image and determined the lens redshift to be $z_{l} \simeq 0.60$.

Oguri et al. (2008b) modeled the positions of the three images using a single elliptical Navarro, Frenk, \& White (NFW; 1997) density profile centered near galaxy G1, consistent with the cluster center inferred from weak lensing and the presence of additional lensed arcs. These models predicted flux ratios of A : B : C = 0.11 : 1.00 : 0.99 that are wildly at odds with the observed optical flux ratios of 0.95 : 1:00 : 0.24. Simply tweaking the parameters of a cluster potential cannot fix this problem because B and C are a fold pair on opposite sides of a critical curve -- in a smooth potential, these components should be more magnified than image A and have a flux ratio of order unity. The optical flux ratios alone cannot resolve this issue because they are influenced by so many physical effects: lensing, microlensing, and extinction.

In this letter we present the results from new radio observations of the system. Emission at these wavelengths is unaffected by dust or stellar microlensing, leaving only substructure as a potential explanation if the anomalies persist.  As we discuss in \S2, we successfully resolved all 3 images and find that anomalies persist but differ from those in the optical. In \S3 we model and interpret these results, and in \S4 we discuss their implications.

\begin{figure*}
\plotone{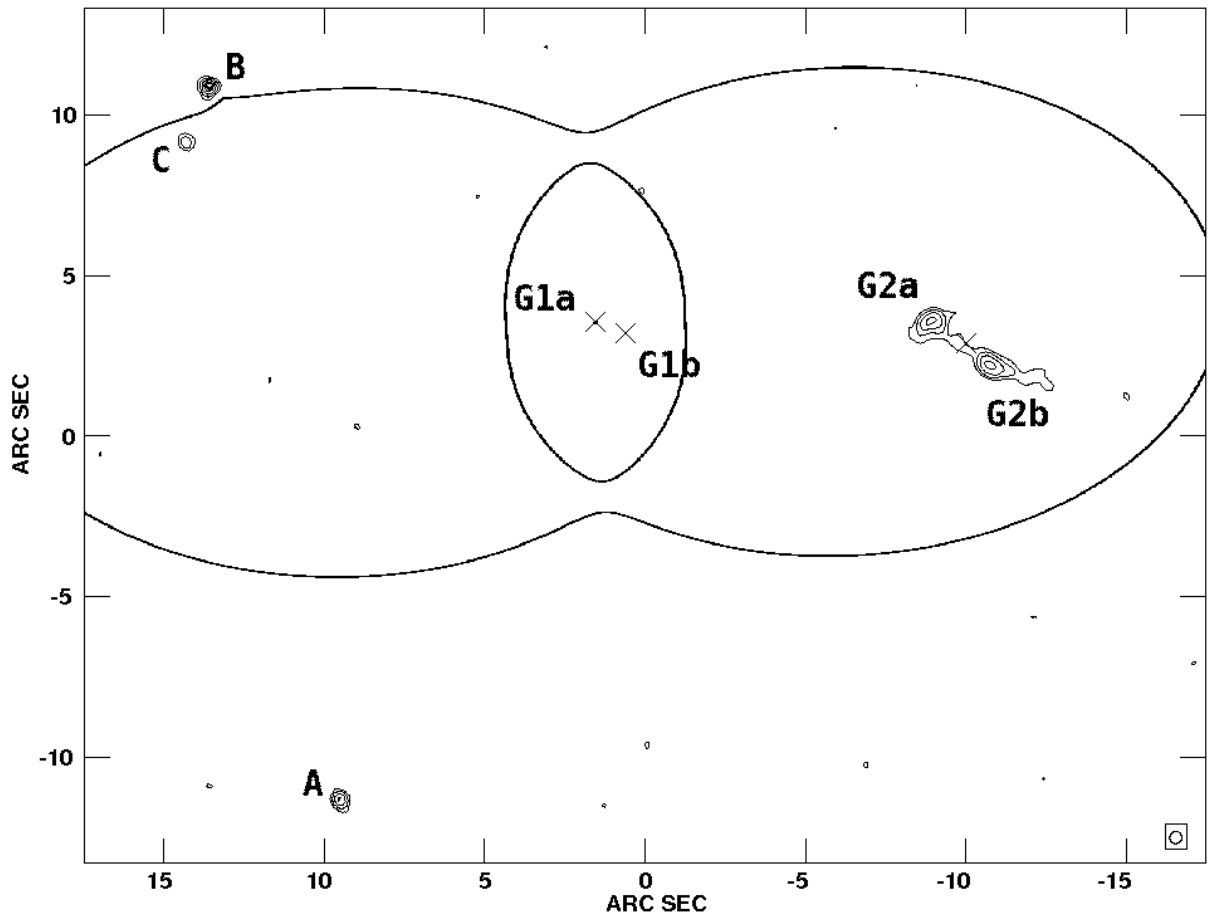} 
\caption{The follow-up 6 cm radio map of SDSS J1029+2623. North is up and East is left. The clean beam shape is shown in the lower right corner. Note that this map resolves images B and C while also establishing that the radio source associated with galaxy G2 is made up of two distinct components, which we label G2a and G2b. The optical center of G2 is marked with a cross and is located at R.A. 10 29 12.49 and Dec. +26 23 32.12 (8.2 m Subaru observations). The solid line indicates the critical curves for our best fit mass model discussed in \S3.1.\label{fig1}}
\end{figure*}

\section{Follow-Up Radio Observations}
We obtained a deep radio map of SDSS J1029+2623 using the NRAO\footnote[9]{The National Radio Astronomy Observatory is operated by Associated Universities, Inc., under cooperative agreement with the National Science Foundation} Very Large Array (VLA) in its A array (the highest resolution configuration). On 2008 November 17, we took thirteen 45 minute C-band (6 cm) observations. For our phase and flux calibrators we used 1018+357 and 1331+305 (3C286; Baars et al. 1977), respectively. The data were reduced and analyzed using standard AIPS tasks.

Our radio map, with an angular resolution $\sim 0\farcs4$, is shown in Figure 1, while the photometry and astrometry are summarized in Table 1. As expected, both the B and C components are seen in the radio map. In our previous 6 cm map, taken with the VLA in the transitional BnC array, they were blended together because of the extended beam associated with the configuration.  In the new map, the beam is more compact and almost circular (see Figure 1), so the radio images were not extended in any preferred direction. This higher resolution radio image also reveals that the radio source associated with galaxy G2 ($\sim 10\arcsec$ West of the main lens galaxy G1) is resolved into two components, which we name G2a and G2b. These appear to be two radio lobes associated with an AGN in G2.

\begin{deluxetable}{rccrrccc}
\tabletypesize{\small}
\tablecaption{SDSS J1029+2623: Radio Astrometry \& Photometry}
\tablewidth{0pt}
\tablehead{
\colhead{Component} & \colhead{RA} & \colhead{Dec.} &\colhead{$\Delta$x (\arcsec)} & \colhead{$\Delta$y (\arcsec)} & \colhead{Peak Flux Density} & \colhead{Integrated Flux}\\
\colhead{} & \colhead{} & \colhead{} & \colhead{} & \colhead{} & \colhead{(mJy/beam)} & \colhead{(mJy)}
}
\startdata
A & 10 29 13.95 & +26 23 17.96 & 0 & 0 & 0.206 & 0.225\\
B & 10 29 14.25 & +26 23 40.16 & $-$4.08 & 22.20 & 0.257 & 0.289\\
C & 10 29 14.30 & +26 23 38.36 & $-$4.80 & 20.40 & 0.117 & 0.129\\
G2a & 10 29 12.57 & +26 23 32.84 & 18.48 & 14.88 & 0.211 & 0.629\\
G2b & 10 29 12.44 & +26 23 31.40 & 20.28 & 13.44 & 0.216 & 0.963
\enddata
\tablecomments{All positions are measured as the position of peak flux density relative to image A. No radio components of G1 were detected. Positional errors are typically $\la 0\farcs05$ while the flux density errors are $\approx$ 0.013 mJy/beam.}
\end{deluxetable}

As far as flux ratios are concerned, the new radio flux ratios of 0.80 and 0.46 for A/B and C/B, respectively, not only fail to match the ratios of 0.10 and 0.99 predicted by smooth lens models (Oguri et al. 2008b), but they also differ from the optical flux ratios measured in a broad range of bandpasses between 2006 and 2008 (see Table 2). In order for the optical and radio flux ratios to differ both with themselves and with the models, there must be multiple physical effects at work.  Explaining the radio flux ratios requires adding substructure to the lens potential, while the optical flux ratios must be further affected by extinction, microlensing, or some other physical effect.

\begin{deluxetable}{rlccclll}
\tablecolumns{7}
\tablewidth{0pt}
\tablecaption{SDSS J1029+2623: Optical \& Radio Flux Ratios}
\tablehead{
\colhead{} & \multicolumn{3}{c}{Component} & \colhead{} & \multicolumn{2}{c}{Flux Ratios}\\
\cline{2-4} \cline{6-7}\\
\colhead{Bandpass} & \colhead{A} & \colhead{B} & \colhead{C} & \colhead{} & \colhead{A/B} & \colhead{C/B}}
\startdata
Optical B & 19.20 & 19.03 & 20.89 & & 0.86 & 0.18\\
        V & 18.72  & 18.67 & 20.63 & & 0.96 & 0.16\\
        g & 18.72  & 18.81 & 20.87 & & 1.09 & 0.15\\
        R & 18.46 & 18.58 & 20.38 & & 1.12 & 0.19\\
        I & 18.01 & 17.95 & 19.51 & & 0.95 & 0.24\\
& & & & & & \\
Radio 6 cm & 0.206 & 0.257 & 0.117 & & 0.802 $\pm$ 0.063 & 0.455 $\pm$ 0.054\\
& & & & & & \\
Smooth Naked & & &  & & 0.12 & 0.99\\
Cusp Model & & &  & &  &
\enddata
\tablecomments{The optical component values are reported as magnitudes (reproduced from Oguri et al. (2008b)) and have errors $\leq$0.02 mag. The optical ratios are reported as fluxes with errors $<$0.03. The radio component values and errors ($\sim$0.013 mJy/beam) are reported as peak flux densities (mJy/beam). Optical monitoring of the data over the last few years shows some intrinsic variability, but nothing on the scale of the B versus A/C flux ratios.}
\end{deluxetable}

\section{Interpretation of the Data}
We continue to model our system as a naked cusp since our radio maps only show three quasar images and all other possible lensing configurations are highly unlikely (Oguri et al. 2008b). Before adding a substructure component, we first reconfirmed the results of Oguri et al. (2008b) using the \emph{lensmodel} package (Keeton 2001) to fit a mass model to our three observed radio positions. Following Oguri et al. (2008b), we modeled the system using an elliptical NFW density profile at the position of G1a (R.A. 10 29 13.35, Dec. +26 23 32.80) reported by Oguri et al. (2008b), since G1 was undetected in our radio observations. We also fixed the scale radius, $r_s$, to 60$\arcsec$ just as Oguri et. al. (2008b), since the scale radius and mass are usually degenerate parameters (Oguri et al. 2004). The radio model closely mirrors the previously reported model, having ellipticity $e = 0.42$ and  position angle $\theta_e = -87.5\degr$  East of North corresponding to the G1-G2 axis (Oguri et al. 2008b). 

Even though the model correctly reproduces the observed radio positions, it again predicts flux ratios of 0.12 and 0.99 for A/B and C/B, respectively, that disagree with the observed 6 cm flux ratios of 0.80 and 0.46. Since we have reconfirmed that a smooth mass model cannot create the observed flux ratios we next explore adding substructure near images B and C.

\subsection{Substructure Parameter Study}
We modeled the substructure as a single isothermal sphere (SIS) added to the NFW model used to fit the radio positions. Using a technique similar to that of MacLeod et al. (2009), we varied the position of the substructure within a $6\arcsec \times 8\arcsec$ region centered on the position of image B. Using the \emph{lensmodel} package, we placed the substructure on a grid in RA and Declination with positions spaced by 0\farcs3. Figure 2 shows the resulting $\chi^2$ surface for the fits after optimizing the mass of the substructure and the parameters of the NFW model.


Placing the substructure Northwest of component B produced the best results, with the substructure roughly $0\farcs1$ West and $0\farcs1$ North of image B. The NFW parameters are little changed ($e = 0.42$ and $\theta_e = -87.5\degr$ East of North) because the required deflection scale of the substructure is very small ($\theta_E=0\farcs011$). The model has a $\chi^2$ per degree of freedom of $0.34$. Figure 1 shows the critical curves of the best fit mass model.

Truncating the isothermal sphere at the approximate projected tidal limit between the cluster and our SIS  perturber ($R_{tidal} = 1.41$ kpc) yields a substructure with a mass $\sim 10^{9}$ M$_\sun$ up to the mass-sheet degeneracy. If, instead, we integrated over the surface mass density up to the same radius, we would get a result that is only larger by a factor of $\pi/2$. This mass estimate would also hold within the order of magnitude given if the substructure were modeled with an NFW density profile. We estimated the effect moderately resolved quasar components would have by using the upper limit (integrated) flux values to calculate the Einstein radius of an SIS located at the same position as our best fit substructure and found no effect on the substructure's Einstein radius or mass estimate. A choice of scale radius for the NFW mass model less than the separation between the NFW and SIS centers ($\sim 22\arcsec$) would reduce this mass estimate, but Oguri et al. (2004) has extensively analyzed the characteristic surface mass density, ${\kappa}_s$, as a function of scale radius, and determined that the only physically likely models are those with $r_s \geq 30\arcsec$; therefore, our mass estimate holds for tenable scale radii.  

\begin{figure}[t]
\plotone{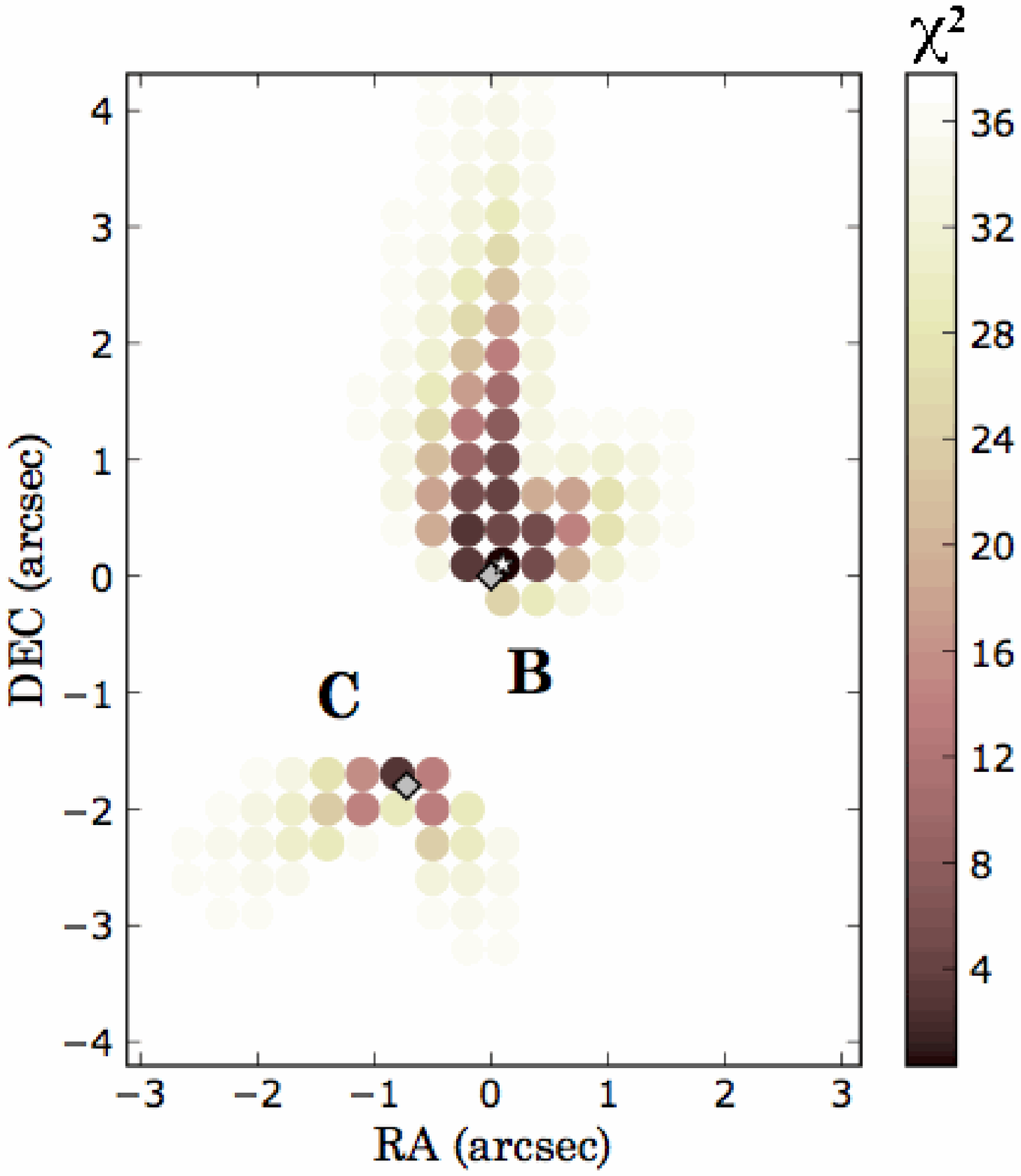} 
\caption{A plot of the $\chi^{2}$ surface as a function of SIS substructure position for three degrees of freedom. The gradient of dots from lighter to darker represents substructure locations that, when combined with our previously reported NFW profile centered on G1a, reproduce our radio observations with increasing accuracy. Lensing components B and C are denoted by the grey diamonds. The substructure position with the smallest $\chi^{2}$ per degree of freedom is denoted with a tiny white star.\label{fig2}}
\end{figure}

All three predicted image positions lie within the errorbars of the measured image positions, while the flux ratios predicted by this model are 0.05 and 0.46 for A/B and C/B, respectively. We consider this a success since we did not take the relative flux of component A into account when computing our models. We have already shown with components B and C that adding a small perturbation near an image to a smoothly varying central potential will significantly change the relative flux of that image without appreciably changing its position (Goldberg et al., 2010). We could simply add more substructure near image A to change its flux without changing the flux ratio between B and C significantly since component A is so far separated from the other image pair. 

There are many degeneracies associated with modeling lensed systems, so our proposed configuration is not unique, but it provides a first step in understanding this unusual system.
 
\subsection{Separation of Optical and Radio Components}

Finally, the difference between the optical and radio flux ratios needs to be accounted for. The optical flux ratios lack the strong wavelength dependence needed to explain the differences with extinction (e.g. see the survey of extinctions in lenses by Falco et al. 1999). To generate the V band flux ratio from the radio flux ratio with dust requires $E(B-V) \simeq 0.6$ mag, which would lead to a large, and unobserved, change in the flux ratio between the B and V bands of almost a factor of two. Microlensing is possible in a cluster environment, and it seems to be observed in the other cluster quasar lens SDSS J1004+4112 (Richards et al. 2004; Fohlmeister et al. 2007). This is somewhat puzzling because the expected surface density in stars is very low. This would best be confirmed by detecting the uncorrelated time variability between the images that is characteristic of microlensing.

There is, however, one additional possibility, namely that the positions of the optical and radio sources are slightly different so that they experience different magnifications from the substructure. In particular, we note that the radio source is a steep spectrum source with $\alpha_\nu \simeq -1.14$, suggesting that it is dominated by unresolved emission from extended radio lobes/jets rather than a compact core.  Using our best fit substructure model and \emph{lensmodel}, we determined how the B/C flux ratio changes as we move the optical source further from the caustic than the radio source. Our radio source resides $\sim$20 mas from the caustic, and we placed `optical sources' in 1 mas increments from the radio source such that each `optical source' receded from the caustic by an additional $\sim$0.1 mas.  Figure 3 shows that as a source gets further from the caustic the flux ratio shrinks.

Even though the optical emission of a quasar is produced in its central engine, we chose to move the `optical source' because our radio source positions are much more accurate than our optical positions. From our analysis, we find that positional offsets of the two regimes on scales smaller than the errors can cause greatly varying flux ratios. Thus, SDSS J1029+2623 could be comprised of a radio jet closer to the caustic with an optical core offset by about 8 - 9 mas. This would not be the first lens whose appearance is affected by an offset between the radio and optical emission. MG 2016+112 is a still more dramatic example, where the optical quasar lies in a two-image region while the radio and X-ray jet components cross into the four image region leading to dramatically different lens morphologies (see Kochanek, Schneider, \& Wambsganss 2004).


\begin{figure}[h]
\plotone{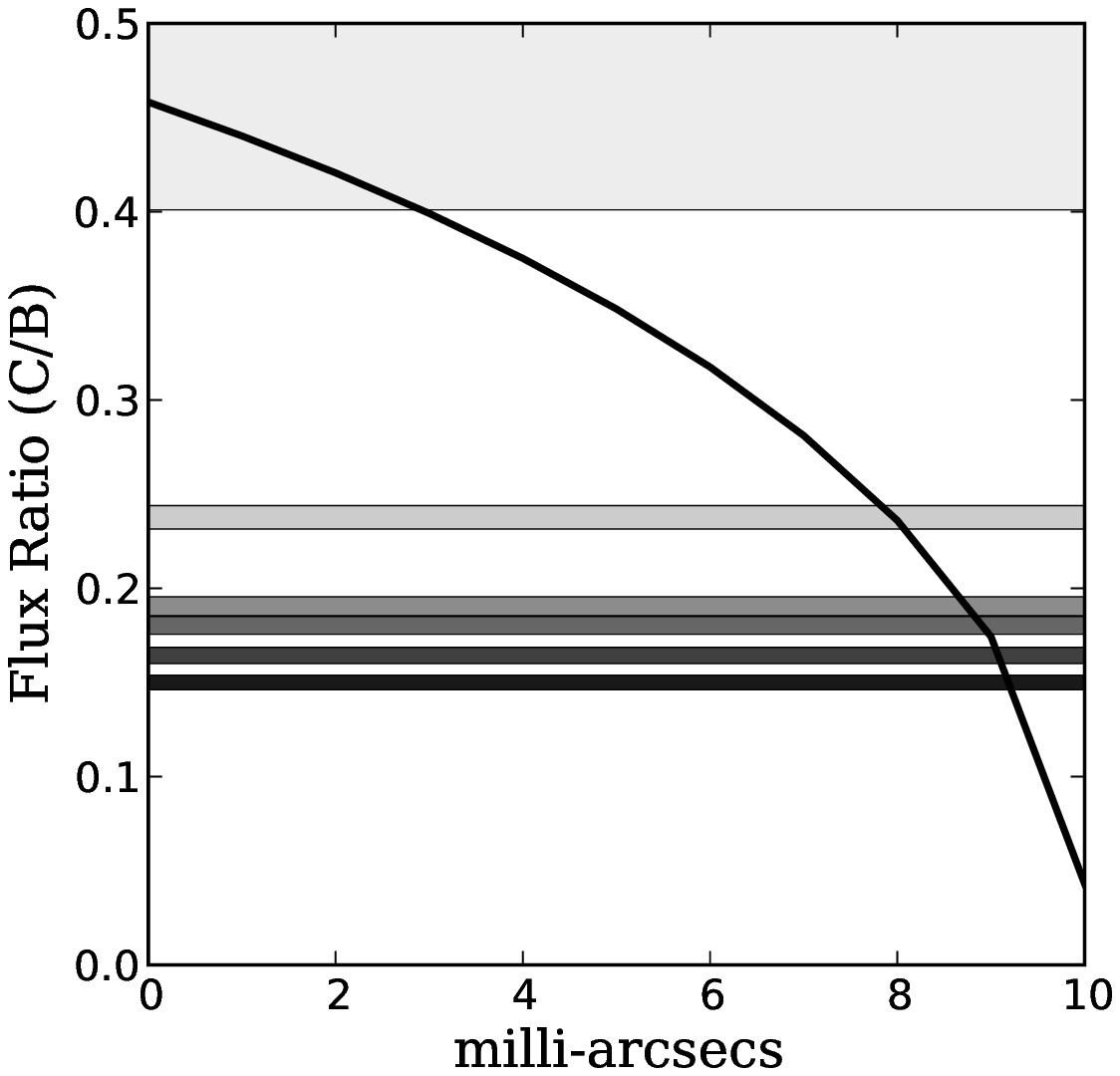}
\caption{The flux ratio between components C and B as a function of radial distance from the radio source position as an `optical source' is moved further from the caustic. The different colored horizontal bands represent the ratio observed in different bandpasses (from top to bottom: 6 cm radio, I, R, B, V, G) (see Table 2). It is evident that as the `optical source' moves away from the caustic, the flux ratio between B and C shrinks.\label{fig3}}
\end{figure}

\section{Conclusions}

The most probable configuration of SDSS J1029+2623 is a quasar at $z_{s} = 2.197$ triply imaged by a cluster of galaxies at $z_{l} \simeq 0.60$ and a dark matter clump with a mass $\sim 10^{9}$ M$_\sun$ slightly offset from the position of image B. From our analysis, we also have found that offset optical and radio emission regions caused by extended radio jets are a probable explanation for the disagreement between the optical and radio flux ratios. Higher resolution radio data acquired with VLBI could possibly resolve the radio jets of SDSS J1029+2623 (see More et al. 2009, Figure 1 vs. Figure 4), although an extremely long exposure time would be necessary.

Additionally, 60 ksec ACIS-S observations made with Chandra on 2010 March 11 detected all three quasar images as well as the lensing cluster. Conclusions about SDSS J1029+2623's X-ray emission have yet to be made, but we are hoping to compare the X-ray properties with our independent strong lensing constraints and definitively eliminate microlensing and dust extinction as possible causes of the flux anomaly by studying the spectrum of the faintest image (Oguri et al., in prep.).

Finally, the Hubble Space Telescope is set to observe SDSS J1029+2623 for 7 orbits in cycle 18, increasing the resolution and depth of our optical data. We plan on searching for faint perturbers near images B and C that may be responsible for the flux anomalies and will try to detect the lensed quasar's host galaxy in order to study the lens potential around components B and C in more detail. Hopefully our multi-bandpass data will firmly establish the origins of this unique lensing configuration.

\acknowledgements

RMK would like to thank Wendy B. Harris for her \emph{lensmodel} expertise. GTR was supported in part by an Alfred P. Sloan Research Fellowship. CSK is supported by NSF grants AST-0708082 and AST-1009756. RHB's work was supported in part under the auspices of the US Department of Energy by Lawrence Livermore National Laboratory under contract W-7405-ENG-48. Based in part on data collected at Subaru Telescope, which is operated by the National Astronomical Observatory of Japan. 

\clearpage

\end{document}